\documentclass[prb,twocolumn,superscriptaddress,showpacs,citeautoscript]{revtex4}
\usepackage{amsmath}
\usepackage{amssymb}
\usepackage{graphicx}


\begin{document}

\title{Spin-dependent thermoelectric effects in a strongly correlated double quantum dot}
\author{\L{}ukasz Karwacki}
\email{karwacki@amu.edu.pl}
\affiliation{Faculty of Physics, Adam Mickiewicz University, 61-614 Pozna\'{n},
Poland}
\author{Piotr Trocha}
\email{ptrocha@amu.edu.pl}
\affiliation{Faculty of Physics, Adam Mickiewicz University, 61-614 Pozna\'{n},
Poland}

\begin{abstract}
We investigate spin-dependent thermoelectric transport through a system of two coupled quantum dots attached to reservoirs of spin-polarized electrons. Generally, we focus on the strongly correlated regime of transport. To this end, a slave-boson method for finite $U$ is employed. Our main goal is to show, that apart from complex low-temperature physics, such basic multi-level system provides a possibility to examine various quantum interference effects, with particular emphasis put on the influence of such phenomena on thermoelectric transport. Apart from the influence of interference effects on spin-degenerate charge transport, we show how spin-dependent transport, induced by ferromagnetic leads, can be modified as well. Finally, we also consider the case, where the spin relaxation time in the ferromagnetic leads is relatively long, which leads to the so-called spin thermoelectric effects.
\end{abstract}

\pacs{73.23.-b, 73.63.Kv\, 72.15.Qm, 85.35.Ds}
\maketitle

\section{Introduction}
The trend in microelectronic fabrication known as Moore's law has recently started to show possible setbacks which can be expected in the near future. Apart from the problems in miniaturization, many other problems result from excessive heat generation. To drive the progress further it is necessary to understand the processes governing heat transport at lower scales and its possible utilization in prospective devices. It is believed that the range of phenomena known as thermoelectric effects can ameliorate some of those problems.

Particular interest has been focused on nanoscale systems and materials in which the quantum-mechanical nature of the phenomena lends itself to increase thermoelectric efficiency.~\cite{Hicks, Dresselhaus, Mahan-1, Shakouri, Goupil} One of such features, which is known to greatly influence thermoelectric response, is discrete structure of the density of states.~\cite{Mahan-bestthermo} This property can be easily realized and tuned in low-dimensional structures such as quantum dots.~\cite{Beenakker, Kubala,Krawiec1, zimbovskaya}

Inclusion of additional conducting channels as in multilevel quantum dots or multiple single-level quantum dots leads to more complex behavior. For instance, enhancement of interdot Coulomb interaction in otherwise electrically separated dots has been experimentally shown to allow a double quantum dot structure to act as a gate and bias voltage controlled electrical current switch, where transport through one of the dots can be turned off if current passes through the other,~\cite{Chan, Huebel} as evidenced by sign reversal of current noise cross correlations~\cite{McClure}. This property has been later on utilized in nanoscale heat engines controlled electrically, based on two chaotic cavities where conversion of thermal voltage fluctuations to electrical current occurs~\cite{Roche, Sothmann-cavity} and on a system of two quantum dots, where voltage fluctuations in one dot led to rectified electrical current through the other.~\cite{Hartmann}

The feature that distinguishes double quantum dots is, apart from the tunability pertinent to quantum dots in general, a possibility to investigate and utilize interference effects. Allowing strong enough electrical contact between the discrete levels results in formation of the so-called bonding and antibonding states in analogy to the hydrogen molecule. Existence of such molecular states has been verified experimentally for semiconductor double quantum dots and carbon nanotube quantum dots and has been shown to greatly influence transport through those multilevel systems in different coupling regimes.~\cite{Chang, Pioro, Holleitner, Blick}
Furthermore, it has been predicted that the electrons propagating coherently through the dots can interfere both constructively and destructively leading to Dicke~\cite{Dicke, Shahbazyan, Trocha-dicke} or Fano effects~\cite{Fano,Guevara,TrochaFano} resulting in enhanced thermoelectric response.~\cite{Kawakami, Trocha-thermo, Wierzbicki, Lim, Liu, GarciaSuarez, Hershfield,Wojcik}

Necessary miniaturization will inevitably lead to a strongly correlated or Kondo regime in transport, which in quantum dot systems results in enhancement of conductance due to the screening of the on-site electron's spin by the electron cloud from the leads.~\cite{Cronenwett, vdWiel, Grobis} Thermoelectric effects in this transport regime have been predicted and experimentally confirmed to present many interesting features such as, e.g., temperature-dependent sign change of the thermopower or strong violation of the Wiedemann-Franz law.~\cite{Boese, Scheibner, Costi} There exists, however, another possibility for a Kondo effect in two-level quantum dots and in double quantum dot (DQD) systems due to orbital degeneracy, where the orbitals serve as a pseudospin degree of freedom.~\cite{Keller} The interplay between strong correlations and various interference effects significantly modifies transport properties of systems based on double quantum dots.~\cite{Lara, Okazaki, Sasaki, Vernek, DingNahm, Tosi, Krychowski, Krychowski2,Koshibae} Particularly, the interplay of Dicke and Kondo effects has been predicted to greatly increase transport characteristics in quantum dot systems.~\cite{Trocha-dicke, Orellana, karwackiJPCM} and is likely to increase thermoelectric power of the system.~\cite{Koshibae,Azema,Donsa, Zhang}

Recent experimental observation of the spin counterparts of the Seebeck and Peltier effects has lead to a new field known as spin caloritronics.~\cite{Uchida, Bauer, Flipse, Hatami, Jansen} These effects were already confirmed in a range of materials such as magnetic metals, magnetic semiconductors or magnetic insulators. However, it was shown theoretically that the intrinsic properties of quantum dot systems that influence charge-dependent effects could affect spin-dependent thermoelectric effects as well.~\cite{Trocha-thermo, Krawiec, Swirkowicz, Dubi, Weymann, Rejec, karwackiJPCM, Sothmann-magnon} One of the particularly characteristic features is the strong dependence of spin thermopower on magnetic polarization of the adjacent leads. In the Kondo regime this effect is additionally enhanced due to the presence of an exchange field which tends to split the Kondo resonance peak.~\cite{Martinek,karwackiJPCM,Wojcik2}

\begin{figure}
\includegraphics[width=\columnwidth]{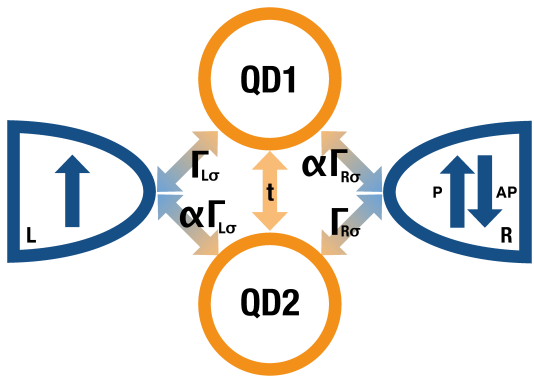}
\caption{\label{Fig:1}The system investigated in the paper consists of two single-level quantum
dots each with Coulomb correlations $U_{i}$ coupled to each other with hopping parameter $t$ and
to magnetic electrode $\beta=L,R$ with the coefficients $(\alpha) \Gamma_{\beta\sigma}$ where $\alpha$ describes possible asymmetry. The configuration of the
leads depends on the value and sign of respective polarizations and
can be either parallel (P) or anti-parallel (AP).}
\end{figure}

The system considered in this paper is presented schematically in Fig.~\ref{Fig:1} and consists of two quantum dots coupled to each other and to two electron reservoirs, which are assumed to be sources of spin-polarized carriers. We show that both charge and spin thermoelectric transport characteristics are significantly modified in the Kondo regime particularly due to interference effects induced by different couplings of the dots' levels to leads, resulting in an electronic analog of Dicke and Fano effects. To this end we employ the slave-boson mean field technique (SBMFT) for finite $U$.~\cite{KotliarRueckenstein, Dong, Lechermann, Fresard, Li} The model and the method are detailed in Sec. II of the paper. We present numerical results in Sec. III, where we distinguish the cases of leads with and without spin accumulation. Additionally, we examine the influence of intrinsic parameters of the system, such as configuration of the dots and leads' spin polarization, on the thermoelectric phenomena. Section IV contains a brief summary of the paper.

\section{Theoretical description}
\subsection{Model and method}
\subsubsection{Double quantum dot system}
The system of two coupled quantum dots
attached to external magnetic electrodes is presented schematically in Fig.~\ref{Fig:1} and can be
described by the following Hamiltonian:
\begin{equation}\label{Eq:1}
\mathrm{H}=\mathrm{H_{e}}+\mathrm{H_{dqd}}+\mathrm{H_{t}}\,.
\end{equation}
The first term, $\mathrm{H_{e}}=\sum_{\mathbf{k}\beta\sigma}\varepsilon_{\mathbf{k}\beta\sigma}c_{\mathbf{k}\beta\sigma}^{\dagger}c_{\mathbf{k}\beta\sigma}$,
describes the Fermi sea of spin-polarized electrons of wave vector
$\textbf{k}$ and spin $\sigma (=\uparrow,\downarrow)$ in the $\beta (=L,R)$th lead. The second term describes the isolated double quantum dot and takes the following form:
\begin{align}\label{Eq:2}
\mathrm{H_{dqd}} &= \sum_{i\sigma}\varepsilon_{i\sigma}d_{i\sigma}^{\dagger}d_{i\sigma}+t\sum_{\sigma}(d_{1\sigma}^{\dagger}d_{2\sigma}+d_{2\sigma}^{\dagger}d_{1\sigma})\nonumber \\
 &+  \sum_{i}U_{i}n_{i\uparrow}n_{i\downarrow}\,,
\end{align}
Here, $\varepsilon_{i\sigma}$
is the energy level of the $i$-th ($i$=1,2) quantum dot, $t$ is the parameter
describing spin-conserving hopping between the dots, whereas $U_i$ denotes the
Coulomb energy corresponding to double occupation of the $i$th dot.
The last term of Hamiltonian (\ref{Eq:1}) describes spin-conserving tunneling between the leads and $i$th dot and takes the form
\begin{equation}\label{Eq:3}
\mathrm{H_{t}}=\sum_\sigma\sum_{\mathbf{k}\beta i}(V_{\beta\sigma i}c_{\mathbf{k}\beta\sigma}^{\dagger}d_{i\sigma}+{\rm H.c.}).
\end{equation}
Here, $V_{\beta\sigma i}$ denotes the relevant matrix element assumed to be independent of wave vector $\mathbf{k}$.
\subsubsection{Slave-boson method}
To describe transport properties of the considered system in the Kondo regime we utilize the slave-boson technique developed for a finite value of the Coulomb parameter $U$, which provides reliable results in the low-temperature regime, excluding possible fluctuations, and also under the condition of spatial uniformity of the spin quantization axis~\cite{KotliarRueckenstein, Lechermann, Li, Fresard}.
In the first step of the method one replaces the initial creation (annihilation) $d_{i\sigma}^{\dagger}$ ($d_{i\sigma}$) operator of each dot by the product of the pseudofermion and boson operators, $f_{i\sigma}^{\dagger}z_{i\sigma}^{\dagger}$ ($z_{i\sigma}f_{i\sigma}$). The operator $z_{i\sigma}^{\dagger}=e_{i}p_{\sigma}^{\dagger}+p_{i\overline{\sigma}}d_{i}^{\dagger}$ ($z_{i\sigma}=p_{i\sigma}e_{i}^{\dagger}+d_{i}p_{i\overline{\sigma}}^{\dagger}$) acts as a projection operator on the extended Fock space of the dot. Each state in this extended space can be defined as $|\chi\rangle_{i} = |b\rangle_{i} \otimes |f\rangle_{i}$, where $|b\rangle_{i}$ is the bosonic and $|f\rangle_{i}$ is the fermionic state of the $i$th dot, respectively. The ground state of the $i$th dot is defined as $|\mathrm{vac}\rangle_{i} = |0\rangle_{i} \otimes |0\rangle_{i}$.

The component operators $e_{i}$, $p_{i\sigma}$, and $d_{i}$ act on empty, singly (with spin $\sigma$) and doubly occupied $i$th dot states, respectively, as follows:
\begin{align}
e_{i}|E\rangle_{i} &= e_{i}|e\rangle_{i} \otimes |0\rangle_{i} = |\mathrm{vac}\rangle_{i}, \nonumber \\
p_{i\sigma}f_{i\sigma}|P_{\sigma}\rangle_{i} &= p_{i\sigma}|p_{\sigma}\rangle_{i} \otimes f_{i\sigma}|\sigma\rangle_{i} = |\mathrm{vac}\rangle_{i}, \nonumber \\
d_{i}f_{i\uparrow}f_{i\downarrow}|D\rangle_{i} &= d_{i}|d\rangle_{i} \otimes f_{i\uparrow}f_{i\downarrow}|\uparrow\downarrow\rangle_{i} = |\mathrm{vac}\rangle_{i}\,.
\end{align}
Accordingly, the product of boson and pseudofermion creation operators acts on the $|\mathrm{vac}\rangle_{i}$ state creating empty, singly (with spin $\sigma$) and doubly occupied states.

However, to ensure physical solutions, one has to introduce necessary boundary conditions, the first of which is the conservation of states $e_{i}^{2}+\sum_{\sigma}p_{i\sigma}^{2}+d_{i}^{2}=1$, whereas the second denotes the correspondence $p_{i\sigma}^{2}+d_{i\sigma}^{2}=\langle f_{i\sigma}^{\dagger}f_{i\sigma} \rangle$. Furthermore, we assume saddle-point approximation for the slave-boson operators which allows for the use of their mean values, $z_{i\sigma}\equiv\langle z_{i\sigma}^{(\dagger)}\rangle$.~\cite{KotliarRueckenstein}

The transformation leads to the effective Hamiltonian of the form:
\begin{equation}\label{Eq:4}
\mathrm{H_{eff}}=E_{sb}+\tilde{\mathrm{H}}\,,
\end{equation}
where the first term assumes the form
\begin{align}\label{Eq:5}
E_{sb}&=\sum_{i}[ U_{i}d_{i}^{2}+\lambda_{i}^{(1)}\left(e_{i}^{2}+\sum_{\sigma}p_{i\sigma}^{2}+d_{i}^{2}-1\right) \\ \nonumber &- \sum_{\sigma}\lambda_{i\sigma}^{(2)}\left(p_{i\sigma}^{2}+d_{i}^{2}\right)]
\end{align}
and represents energy related to double occupancy of the dots along with the aforementioned constraints introduced with Lagrange multipliers $\lambda_{i}^{(1)}$ and $\lambda_{i\sigma}^{(2)}$.
This transformation acts only on the dot operators so the term $\mathrm{H_{e}}$ of the original Hamiltonian (\ref{Eq:1}) remains unmodified. Thus, $\tilde{\mathrm{H}}=\tilde{\mathrm{H}}_\mathrm{dqd} + \tilde{\mathrm{H}}_\mathrm{t} + \mathrm{H_{e}}$. The double quantum dot Hamiltonian now takes the form
\begin{equation}\label{Eq:6}
\tilde{\mathrm{H}}_\mathrm{dqd} =\sum_{i\sigma}\tilde{\varepsilon}_{i\sigma}f_{i\sigma}^{\dagger}f_{i\sigma}+\sum_{\sigma}\tilde{t}_{\sigma}(f_{1\sigma}^{\dagger}f_{2\sigma}+{\rm H.c.})\,.
\end{equation}
Here, $\tilde{\varepsilon}_{i\sigma}=\varepsilon_{i\sigma}+\lambda_{i\sigma}^{(2)}$ and $\tilde{t}_{\sigma}=tz_{1\sigma}z_{2\sigma}$.
The tunneling Hamiltonian (\ref{Eq:3}) can now be written down as $\tilde{\mathrm{H}}_\mathrm{t}=\sum_{\beta\sigma i}(\tilde{V}_{\beta\sigma i}c_{\mathbf{k}\beta\sigma}^{\dagger}f_{i\sigma}+{\rm H.c.})$ with $\tilde{V}_{\beta\sigma i}=z_{i\sigma}V_{\beta\sigma i}$.
This parameter is used later to acquire coupling coefficient $\tilde{\Gamma}_{ij\sigma}=2\pi\tilde{V}_{\beta\sigma i}\tilde{V}_{\beta\sigma j}^{*}\rho_{\beta\sigma}$, where $\rho_{\beta\sigma}$ is the density of states in the electrode $\beta$. We assume that the couplings
are constant within the electron band.

When the interdot hopping is of the order of dot-lead coupling, the so-called bonding and antibonding states are formed. Diagonalization of the Hamiltonian results in energies $\tilde{\varepsilon}_{\eta\sigma}$ of antibonding ($\eta=\mathrm{a}$) or bonding ($\eta=\mathrm{b}$) states, respectively (the plus sign corresponds to $\eta=\mathrm{b}$):~\cite{DiagDQD}
\begin{equation}
\tilde{\varepsilon}_{\eta\sigma} = \frac{1}{2}\left(\tilde{\varepsilon}_{1\sigma}+\tilde{\varepsilon}_{2\sigma}\pm \sqrt{(\tilde{\varepsilon}_{1\sigma}-\tilde{\varepsilon}_{2\sigma})+4\tilde{t}_{\sigma}^{2}}\right)\,.
\end{equation}
Matrix elements $\tilde{V}_{\beta\sigma\eta}$ corresponding to the coupling of antibonding and bonding states to electrode states can now be defined using coupling coefficients as $\tilde{V}_{\beta\sigma\eta}=\sqrt{2}/2(\tilde{V}_{\beta\sigma 2} \pm \tilde{V}_{\beta\sigma 1})$.
Coulomb interaction is taken into consideration effectively in slave-boson parameters.

The unknown parameters, $\lambda_{i}^{(1)}$, $\lambda_{i\sigma}^{(2)}$, $e_{i}$, $p_{i\sigma}$, and $d_{i}$, have to be found self-consistently with the help of equations obtained using the Hellman-Feynman theorem applied to the effective Hamiltonian, i.e., $\partial_{\chi_{i}}\mathrm{H_{eff}}=0$ with $\chi_{i}=$($\lambda_{i}^{(1)}$, $\lambda_{i\sigma}^{(2)}$, $e_{i}$, $p_{i\sigma}$,$d_{i}$) for $i=1,2$.
Along with the conservation equations it leads to the following formulas:
\begin{gather}\label{Eq:7}
e_{i}^{2}+\sum_{\sigma}p_{i\sigma}^{2}+d_{i}^{2}-1 = 0\,, \\
p_{i\sigma}^{2}+d_{i}^{2}-K_{0,i,\sigma} = 0\,, \\
\sum_{\sigma}\partial_{e_{i}}\ln z_{i\sigma}K_{1,i,\sigma}+\lambda_{i}^{(1)}e_{i} = 0\,, \\
\sum_{\sigma}\partial_{p_{i\sigma'}}\ln z_{i\sigma}K_{1,i,\sigma}+\left(\lambda_{i}^{(1)}-\lambda_{i\sigma'}^{(2)}\right)p_{i\sigma'} = 0\,, \\
\sum_{\sigma}\partial_{d_{i}}\ln z_{i\sigma}K_{1,i,\sigma}+\left(U_{i}+\lambda_{i}^{(1)}-\sum_{\sigma}\lambda_{i\sigma}^{(2)}\right)d_{i} = 0\,.
\end{gather}
Here, $K_{k,i,\sigma}=(1/2\pi i)\int d\varepsilon(\varepsilon-\tilde{\varepsilon}_{i\sigma})^{k}G_{ii,\sigma}^{<}$ for $k=0,1$ and
$G_{ii,\sigma}^{<}$ is the Fourier transform of the lesser Green's function defined as, $G_{ii,\sigma}^{<}(t,t')=i\langle f_{i\sigma}^{\dag}(t')f_{i\sigma}(t)\rangle$. The lesser Green's function $\mathbf{G}_{\sigma}^{<}$, with the matrix elements $G_{ii,\sigma}^{<}$, can be derived by applying the relation
$\mathbf{G}_{\sigma}^{<}=i\mathbf{G}_{\sigma}^{r}\left(f_{L}\tilde{\mathbf{\Gamma}}_{L\sigma}+f_{R}\tilde{\mathbf{\Gamma}}_{R\sigma}\right)\mathbf{G}_{\sigma}^{a}$, where $f_{L(R)}$ is the Fermi-Dirac distribution for the left (right) electrode, whereas $\mathbf{G}_{\sigma}^{r}$ and $\mathbf{G}_{\sigma}^{a}=\left(\mathbf{G}_{\sigma}^{r}\right)^{\dagger}$ denote retarded and advanced Green's functions, respectively. Furthermore, $\tilde{\mathbf{\Gamma}}_{L(R)\sigma}$ stands for the effective coupling matrix between the dots and the left (right) electrode.

\subsubsection{Green's functions}
The retarded Green's function matrix can be derived using
Dyson's equation $\mathbf{G}_{\sigma}^{r}=\left(\mathbf{g}_{0\sigma}^{r-1}-\mathbf{\Sigma}_{\sigma}\right)^{-1}$,
where [$\mathbf{g}_{0\sigma}^{r}]_{ij,\sigma}\equiv g_{ij\sigma}^{r}=\delta_{ij}\left(\varepsilon-\tilde{\varepsilon}_{i\sigma}+i0^{+}\right)^{-1}$ and
$\mathbf{\Sigma}_{\sigma}=\mathbf{\Sigma}_{d\sigma} + \sum_{\beta}\mathbf{\Sigma}_{\beta\sigma}$
is the sum of effective dot-dot and lead-dot self-energies. The effective dot-dot self-energy matrix takes the form
\begin{equation}\label{Eq:8}
\mathbf{\Sigma}_{d\sigma}=
\begin{pmatrix}
0 & \tilde{t}_{\sigma}\\
\tilde{t}_{\sigma} & 0
\end{pmatrix}\,,
\end{equation}
 whereas the effective lead-dot self-energies can be related to the coupling coefficients via the formula $\mathbf{\Sigma}_{\beta\sigma}=-(i/2)\tilde{\mathbf{\Gamma}}_{\beta\sigma}$ with the renormalized coupling  matrix of the form
\begin{equation}\label{Eq:8a}
\tilde{\mathbf{\Gamma}}_{\beta\sigma}=\begin{pmatrix}\tilde{\Gamma}_{11\beta\sigma} & q\sqrt{\tilde{\Gamma}_{11\beta\sigma}\tilde{\Gamma}_{22\beta\sigma}}\\
q\sqrt{\tilde{\Gamma}_{11\beta\sigma}\tilde{\Gamma}_{22\beta\sigma}} & \tilde{\Gamma}_{22\beta\sigma}
\end{pmatrix}\,.
\end{equation}
The renormalized elements of the coupling matrix, $\tilde{\mathbf{\Gamma}}_{\beta\sigma}$ can be parametrized in the following way: $\tilde{\Gamma}_{11L\sigma}=z_{1\sigma}^{2}\Gamma_{L\sigma}$ and $\tilde{\Gamma}_{22L\sigma}=\alpha z_{2\sigma}^{2}\Gamma_{L\sigma}$, $\tilde{\Gamma}_{11R\sigma}=\alpha z_{1\sigma}^{2}\Gamma_{R\sigma}$ and $\tilde{\Gamma}_{22R\sigma}=z_{2\sigma}^{2}\Gamma_{R\sigma}$. Here, $\Gamma_{\beta\sigma}=\left(1 + \hat{\sigma}p_{\beta} \right)\Gamma$ with $\hat{\sigma}=1$ ($\hat{\sigma}=-1$) for $\sigma=\uparrow$ ($\sigma=\downarrow)$ and $p_{\beta}$ denotes the polarization of the lead $\beta$. For parallel configuration of the leads $p_{\beta}=p$, whereas for antiparallel configuration $p_{L}=|p_{R}|=p$. The parameter $\alpha \in \langle 0,1 \rangle$ describes the difference in the coupling of a given electrode to
the two dots ($\alpha=0$ corresponds to serial configuration whereas $\alpha=1$ corresponds to symmetric parallel configuration).
Parameter $q$ introduced in  Eq.~(\ref{Eq:8a}) describes the strength of indirect coupling of the dots' states through intermediate states of the leads which leads to various interference effects.~\cite{TrochaFano} This parameter mainly depends on the physical distance between the dots in the system. If the distance is of the order of or much smaller than the wavelength of electrons in the reservoir $\beta$, then $q\rightarrow 1$, whereas $q=0$ when the dots, or more generally conducting channels, are well separated.~\cite{Shahbazyan,kubo1,kubo2}

This parameter is motivated by experimental work done on parallel double quantum dots and the investigation of SU(4) and SU(2) Kondo effects in such systems. Ideally, a $q=0$ is necessary for full realization of the SU(4) effect, however it is difficult experimentally to maintain such perfect separation, thus the influence of interdot mixing on transport characteristics is taken into account.

\subsection{Thermoelectric effects}
To describe thermoelectric effects in transport through the double quantum dot system we consider the linear response regime.~\cite{Callen} For our purpose here the forces are bias voltage $\Delta V$, the difference in temperatures of the leads $\Delta T$ and bias spin voltage $\Delta V_{s}$. The last quantity can emerge in magnetic electrodes due to relatively slow spin relaxation resulting in spin accumulation.~\cite{Hatami, Swirkowicz} Additionally, a spin-dependent temperature difference $\Delta T_{s}$ can be introduced; however, in the case considered in this paper such a term can be neglected due to the fact that the energy relaxation time in the leads is shorter than the spin-relaxation time. Thus, it is reasonable to assume that temperature is independent of spin.
According to our assumptions, the chemical potential and temperature of
the left electrode are $\mu_{L}=\mu + e\Delta V_{\sigma}$ and $T_{L}=T + \Delta T$ , respectively,
whereas those of the right electrode are $\mu_{R}=\mu$ and $T_{R}=T$. Here, $\Delta V_{\sigma}=\Delta V + \hat{\sigma}\Delta V_{s}$. Explicitly, $\Delta V=\Delta V_{\uparrow} + \Delta V_{\downarrow}$, and $\Delta V_{s}=\Delta V_{\uparrow} - \Delta V_{\downarrow}$.  The driven charge, spin and heat currents can be written in the form~\cite{Mahan-book}:
\begin{equation}
\left( \begin{array}{c}
J \\
J_{s}\\
J_{q}
\end{array}\right) = \sum_{\sigma}\left( \begin{array}{ccc}
e^2 L_{0\sigma} & e^2 \hat{\sigma}L_{0\sigma} & \frac{e}{T}L_{1\sigma} \\
\frac{\hbar}{2}e \hat{\sigma}L_{0\sigma} & \frac{\hbar}{2}e L_{0\sigma} & \frac{\hbar}{2}\frac{1}{T}L_{1\sigma} \\
e L_{1\sigma} & e \hat{\sigma}L_{1\sigma} & \frac{1}{T}L_{2\sigma}
\end{array}\right)\left( \begin{array}{c}
\Delta V \\
\Delta V_{s}\\
\Delta T
\end{array}\right)\,.
\end{equation}

Here, $L_{n\sigma}=(1/h)\int d\varepsilon\left(\varepsilon-\mu\right)^{n}\left(-\partial_{\varepsilon} f \right)_{T,\mu}T_{\sigma}(\varepsilon)$ for $n=0,1,2$ is the Onsager transport coefficient which relates currents to the driving forces. $T_{\sigma}(\varepsilon)={\rm Tr}(\mathbf{G}_{\sigma}^{a}\tilde{\mathbf{\Gamma}}_{R\sigma}\mathbf{G}_{\sigma}^{r}\tilde{\mathbf{\Gamma}}_{L\sigma})$ denotes the transmission coefficient of the system for the spin $\sigma$  channel.
One can then introduce transport parameters of the system, the first of which is conductance
\begin{equation}\label{Eq:10}
G = e^{2}\sum_{\sigma}L_{0\sigma}\,.
\end{equation}
Let us first consider the situation when no spin bias is generated, i.e., $\Delta V_{s}=0$. To obtain thermoelectric parameters one has to assume open-circuit conditions i.e., $J=0$, which directly leads to the expression for thermopower:
\begin{equation}\label{Eq:11}
S = -\frac{\Delta V}{\Delta T}=-\frac{1}{|e|T}\frac{\sum_{\sigma}L_{1\sigma}}{\sum_{\sigma}L_{0\sigma}}\,,
\end{equation}
Next is an electronic contribution to heat conductance
\begin{equation}\label{Eq:12}
\kappa = \frac{1}{T}\left[\sum_{\sigma}L_{2\sigma}-\frac{\left(\sum_{\sigma}L_{1\sigma}\right)^{2}}{\sum_{\sigma}L_{0\sigma}} \right]\,.
\end{equation}
In turn, for nonzero spin bias, i.e., $\Delta V_{s}\neq 0$, the thermoelectric coefficients are calculated on the condition of vanishing simultaneously both spin current and
charge current, or equivalently under the condition of vanishing charge current in each spin channel. Under these conditions a spin conductance can be defined as
\begin{equation}\label{Eq:10a}
G_{s} = \frac{e\hbar}{2}\sum_{\sigma}\hat{\sigma}L_{0\sigma}\,.
\end{equation}
Next, thermopower assumes the following form
\begin{equation}\label{Eq:11a}
S = -\frac{\Delta V}{\Delta T}=\frac{1}{2}\sum_{\sigma}S_{\sigma}\,,
\end{equation}
where spin $\sigma$ contribution to thermopower can be expressed as follows:
\begin{equation}
S_{\sigma}=-\frac{1}{|e|T}\frac{L_{1\sigma}}{L_{0\sigma}}\,.
\end{equation}
and additionally the so-called spin thermopower can be defined as follows
\begin{equation}\label{Eq:14}
S_{s} = -\frac{\Delta V_{s}}{\Delta T}=\frac{1}{2}\sum_{\sigma}\hat{\sigma}S_{\sigma}\,.
\end{equation}
Electronic contribution to heat conductance takes now the form
\begin{equation}\label{Eq:13}
\kappa = \frac{1}{T}\sum_{\sigma}\left(L_{2\sigma}-\frac{L_{1\sigma}^{2}}{L_{0\sigma}} \right)\,.
\end{equation}
One should bear in mind that for the case  of nonzero spin bias the Fermi distribution function in $L_{n\sigma}$ is now spin dependent.
Furthermore, to remark on the system's thermoelectric efficiency, dimensionless coefficients such as figure of merit and spin figure of merit $Z_{(s)}T=|G_{(s)}|S_{(s)}^{2}T/\kappa$ are introduced.

Now we would like to present numerical results. The corresponding section is divided into three parts, where in the first section we consider a special case of thermoelectric effects in a nonmagnetic system ($p=0$), in the second section we consider thermoelectric effects in a magnetic system ($p\neq0$) without spin bias, and in the last section the case of spin thermoelectric effects is taken into account. Additionally, throughout the sections we consider, among others, two distinct cases of symmetrical coupling of the dots to the leads ($\alpha=1$) and of asymmetrical coupling ($\alpha\neq 1$).

In the following numerical calculations we assume
spin-degenerate and equal dots' energy levels, $\varepsilon_{i\sigma}=\varepsilon_d$ for $i=1,2$ and $\sigma=\uparrow, \downarrow$ ($\varepsilon_d$ is measured
from the Fermi level of the leads in equilibrium,
$\mu_L = \mu_R = 0$). Moreover, we assume $\Gamma=1$ as the
energy unit and the bandwidth is assumed to be
$D=60\Gamma$. Furthermore, the onsite Hubbard parameter is assumed to be the same on each dot, $U_1=U_2\equiv U$, and is equal to $U=6\Gamma$. The dots' energy levels ($\varepsilon_d$) depicted in figures are presented in relation to the parameter $U$.

\section{Nonmagnetic leads}
To clarify strongly correlated physics, we investigate first the case with non-polarized leads ($p=0$). We have calculated such basic thermoelectric coefficients as thermopower $S$, heat conductance $\kappa$, charge conductance $G$, and corresponding figure of merit $ZT$. To deal with various coherent effects, we consider different temperatures, as well as their effect on conductance and thermopower. However, firstly we would like to address the underlying physics of bonding and antibonding states that plays an important role in improving thermoelectric efficiency of the system.

\subsection{Interference effects}
\begin{figure}
\includegraphics[width=\columnwidth]{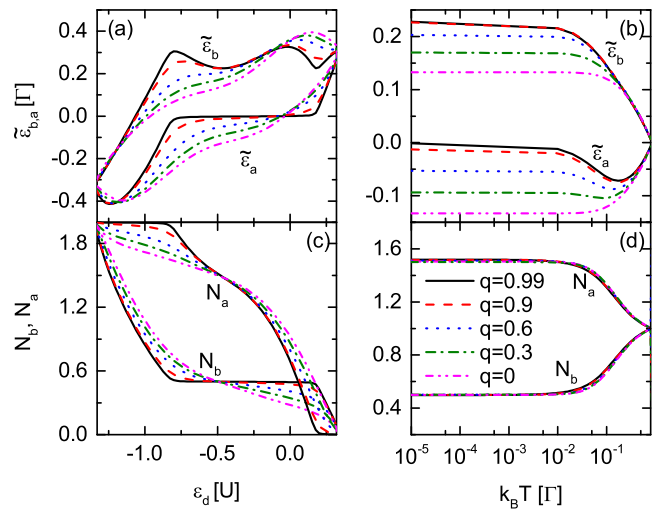}
\caption{\label{Fig:2} Energy $\tilde{\varepsilon}_{b,a}$ of the bonding and antibonding state as a function of dots' energy level, (a), and temperature, (b), and occupation of the bonding and antibonding state as a function of dots' energy level, (c), and temperature, (d), for indicated values of parameter $q$. The other parameters: $U=6\Gamma$, $\alpha=1$, $t=1\Gamma$, $k_{B}T = 0.001\Gamma$ [(a) and (c)], and $\varepsilon_{d}=-U/2$ [(b) and (d)]. In (a) and (b) topmost curves correspond to the bonding state, while in (c) and (d) bottommost curves correspond to the bonding state.}
\end{figure}

For the case of the dots indirectly coupled to each other, transmission through the system can be described by the following formula:
\begin{equation}\label{eq:TransDicke}
T(\varepsilon)=\frac{\tilde{\Gamma}_{b}^2}{(\varepsilon-\tilde{\varepsilon}_{b})^2+\tilde{\Gamma}_{b}^2}+\frac{\tilde{\Gamma}_{a}^2}{(\varepsilon-\tilde{\varepsilon}_{a})^2+\tilde{\Gamma}_{a}^2}\,,
\end{equation}
where $\tilde{\varepsilon}_{b(a)}=\tilde{\varepsilon}_{d} \pm \tilde{t}$ and $\tilde{\Gamma}_{b(a)}=(1\pm q)\tilde{\Gamma}$. Although this formula acquires one-particle transmission form, SBMFT effectively describes the many-body problem.

The superradiant state is a fast decaying state that allows for the virtual exchange of electrons between the dots and electrodes. When the intermediate coupling is maximal, i.e., $q=1$, the subradiant state is no longer active, leaving only  the $2\tilde{\Gamma}$-broadened superradiant state.
Fig.~\ref{Fig:2} shows the dependence of the bonding and antibonding state energies $\tilde{\varepsilon}_{b,a}$ and the occupation of respective states on the dots' level position and temperature for indicated values of the parameter $q$.

Analysis of Fig.~\ref{Fig:2}(a) indicates that for $q=0$ a hopping-induced pseudo-Zeeman splitting occurs with a gap $\tilde{\varepsilon}_{b}-\tilde{\varepsilon}_{a}=2\tilde{t}$, except for the values of dots' energy corresponding to the empty and doubly occupied dots. It is important to note that the initial hopping parameter $t$ is renormalized due to the coupling between dots and leads, thus its effective value is lower than the Kondo temperature of the system (see Sec.~\ref{sec:temp}).  Important resonances occur, when $\tilde{\varepsilon}_{b}=0$ around $\varepsilon_{d}=-U$ and  for $\tilde{\varepsilon}_{a}=0$ around $\varepsilon_{d}=0$.

With increase in $q$, some asymmetry between the states appears due to increasing weight of the bonding state and narrowing weight of the antibonding state. For $q=0.99$ the antibonding state's energy remains constant at the Fermi level ($\tilde{\varepsilon}_{a}=0$) for a broad range of the dots' energy  and allows for the Kondo effect to occur. It is clear that the interplay of interference effects and Kondo physics plays a crucial role in sustaining transport. In the Coulomb regime a Fano effect is more pronounced, with the conductance exhibiting antiresonance for dots' energy corresponding to the antibonding state. ~\cite{Guevara}

Figure~\ref{Fig:2}(b) shows temperature dependence of the bonding and antibonding state's energy for $\varepsilon_{d}=-U/2$. For $k_{B}T\leq 0.01\Gamma$ values of the respective energies saturate. For higher temperatures, i.e., $k_{B}T\geq 1\Gamma$, where a Coulomb blockade regime is dominant and the SBMFT no longer applies, energies of the respective states are equal, as the thermal energy is greater than the energy difference between molecular  levels. For $q=0$, energy of the bonding and antibonding states is symmetric with respect to the $\tilde{\varepsilon}_{b,a}=0$ line (Fermi level). Increase in $q$ leads to a decrease in the antibonding state's energy, effectively pinning it to the Fermi level, as described in the previous paragraph, while the bonding state's energy is shifted away from the Fermi level. In an intermediate temperature range, however, especially for $k_{B}T \approx k_{B}T_{K} \approx 0.25\Gamma$ (calculated as a temperature for which the corresponding conductance reaches half of its maximum value), energy of the antibonding state has a minimum, marking a crossover between the Coulomb blockade and strongly-correlated regimes.

Apart from analyzing the energy landscape of the system, it is nonetheless important to investigate electron occupation. Occupation of both bonding and antibonding states is shown in Fig.~\ref{Fig:2}(c) and (d) as a function of dots' energy level and temperature, respectively. Fig.~\ref{Fig:2}(c) indicates that the electron wavefunction is smeared across the states, with higher occupation of the antibonding state. For increasing $q$ and $\varepsilon_d\approx 0$, occupation of the antibonding state slightly decreases, while that of the bonding state increases. As $q \geq 0.9$, occupation of the bonding state becomes constant for a wide range of dots' energy, while occupation of the antibonding state increases. This effect is a direct result of narrowing weight of the antibonding state and increasing weight of the bonding state. Considering the lifetime of the electron state $\tau_{b,a}=1/\tilde{\Gamma}_{b,a}$ it is clear that the bonding state is characterized by a lower lifetime than the antibonding state.

Another interesting feature is roughly constant occupation of both states in the deep Kondo regime ($\varepsilon_{d}=-U/2$), which has been additionally shown in Fig.~\ref{Fig:2}(d) as a function of temperature. Here occupation of both states, as has been previously stated, is effectively independent of changes in parameter $q$.

The above analysis of the interference-induced effects in the system allows us now to focus on their influence on thermoelectric transport.

\subsection{Case of symmetrical coupling, ($k_{B}T/\Gamma = 0.01$)}
\begin{figure}
\includegraphics[width=\columnwidth]{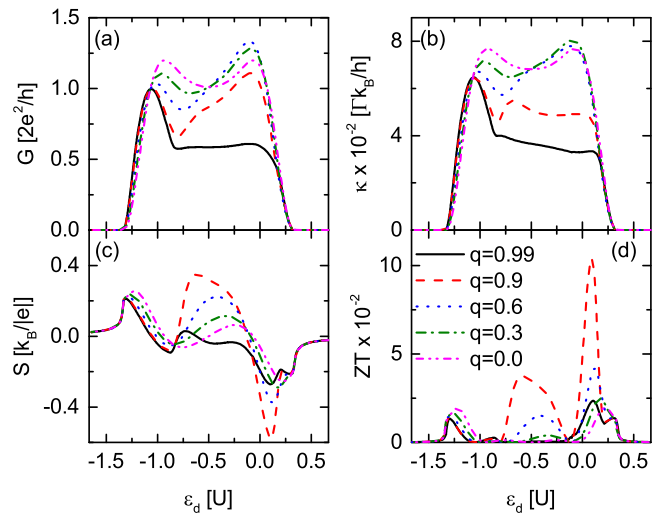}
\caption{\label{Fig:3} Electrical conductance $G$ (a), heat conductance $\kappa$ (b), thermopower $S$ (c), and figure of merit $ZT$ (d), calculated as a function of
the dots' energy level for indicated values of parameter $q$. The other parameters: $k_{B}T = 0.01\Gamma$, $U=6\Gamma$, $\alpha=1$, $t=1\Gamma$.}
\end{figure}

In Fig.~\ref{Fig:3} we present the influence of the indirect coupling strength, measured by the  parameter $q$, on the thermoelectric characteristics. When there is no indirect coupling, $q=0$, the conductance reveals two maxima related to
the resonances at $\varepsilon_d\approx 0$ and $-U$. In the particle-hole symmetry point, $\varepsilon_d=-U/2$, the conductance achieves the minimum. The conductance does not reach the quantum limit due to splitting of the Kondo peak caused by finite hopping between the dots, which lifts the dots' degeneracy.
It has been shown that in strongly correlated two-level systems with both spin and orbital degeneracy, electrical conductance can reach the maximal value of $4e^2/h$. Breaking of the degeneracy, here the orbital one, by allowing electrical contact between the dots ($t>0$), results in smaller values of maximal conductance.
For sufficiently large $t$, the local spin singlet becomes formed suppressing  the  Kondo feature. Then, the conductance exhibits two Hubbard peaks with a deep valley between them (not shown) and maximal values of the conductance reaching $2e^2/h$.

Furthermore, when the indirect tunneling between the dots is on, the bonding (superradiant) and antibonding (subradiant) states become formed. As a result strong asymmetry occurs in the conductance as the parameter $q$ increases. Moreover, for $q\approx 1$, the conductance achieves a maximal value equal to $2e^2/h$ when the peak in the density of states related with the bonding state is in resonance with the Fermi level.
On the other hand, when the dots are in the singly occupied regime, a flat region appears in the conductance. The conductance at the plateau is suppressed to $e^2/h$. However, the conductance in the plateau is not universal and is rather determined by the ratio of $t/\Gamma$\cite{DingNahm}.
The existence of the plateau for the gate voltages corresponding to the singly occupied dots can be explained by gate voltage dependence of the spectral density \cite{Vernek}.
In contrast to the $q=1$ case~\cite{DingNahm,Vernek}, where the antibonding state is decoupled and the only contribution comes from the bonding state, the antibonding state for $q\approx 1$ has a finite coupling to the leads' states and also contributes significantly to transport.

The total contribution (originated from both peaks) to the electronic transport becomes roughly constant in the broad range of the dots' energy, resulting in a plateau in the conductance.
With further decreasing of $\varepsilon_d$, the antibonding state quickly moves away from the Fermi level and the main contribution to the conductance arises from the bonding state which becomes centered at the Fermi level.
%

The $q$ dependence of the electronic contribution to the heat conductance, shown in Fig. ~\ref{Fig:3}(b), follows the dependence of the electron conductance in Fig.~\ref{Fig:3}(a). This is consistent with other low-temperature results~\cite{karwackiJPCM}.
The Seebeck coefficient, shown in Fig.~\ref{Fig:3}(c), changes sign when dots' energy level $\varepsilon_d$ crosses one of the relevant resonances. For $q=0$, the thermopower vanishes for $\varepsilon_d\approx -\tilde{t}$ and $U-\tilde{t}$. The Seebeck coefficient is also zero  in the electron-hole symmetry point $\varepsilon_d=-U/2$. The vanishing of the thermopower in these points is associated with the compensation of charge current due to electrons by that due to holes. With increasing parameter $q$, the behavior of thermopower becomes more complex.

First, the Seebeck coefficient does not vanish in the particle-hole symmetry point for $q>0$. The thermopower calculated for $q=0$ is zero at the symmetric point because the dots' density of states is perfectly symmetric with respect to the energy $\varepsilon=0$. However, with increasing $q$, the density of states becomes strongly asymmetric with respect to the zero energy point, which results in a finite thermopower.
Moreover, the points where thermopower vanishes change their positions. For $q=0.6$, one can notice that $S$ vanishes for $\varepsilon_d\approx -0.1U$ when the antibonding state is in resonance with the Fermi level. The second point in which the thermopower changes sign corresponds to $\varepsilon_d\approx -U$, when the bonding state is in resonance with the Fermi level.
One more point is situated in the valley between the broad and narrow resonances as a result of ``local'' bipolar effect~\cite{Trocha-thermo}.
Then, the current due to electron tunneling through the bonding state is compensated by the
current due to holes tunneling through the antibonding level.
A relatively large value of the thermopower can be found in the vicinity of the antibonding state when $q$ increases. However, for $q$ close to unity the situation becomes more complex and the Seebeck coefficient reveals quantitative and qualitative differences.

The thermopower calculated for $q=0.99$ becomes suppressed for level positions corresponding to the flat region in the conductance.
Despite of the plateau in the conductance, the thermopower shows nontrivial dependence as the dots' level is sweeping. For $\varepsilon_d/U>-0.25$, both peaks in the density of states (bonding and antibonding) are situated slightly above the Fermi level. Thus, through both peaks only electrons can be transmitted, which results in the negative sign of the Seebeck coefficient. With the level position $\varepsilon_d$ decreasing below the value $\varepsilon_d/U=-0.25$, the peak associated with the antibonding state passes beyond the Fermi level and the current due to holes becomes activated resulting in a reduction of the absolute value of the thermopower. For a certain dots' level position, the thermopower vanishes due to the ``local'' bipolar effect mentioned above. With further decreasing $\varepsilon_d$ the Seebeck coefficient acquires positive values.

Fig.~\ref{Fig:3}(d) shows the dots' energy dependence of the figure of merit $ZT$. The best thermoelectric efficiency can be found for large $q$ (but not close to unity). In this case, $ZT$ achieves the largest value in the Coulomb blockade region and when the antibonding level is close to the Fermi level of the leads. In turn, for $q$ close to unity the figure of merit becomes suppressed and its dependence is similar to that calculated for $q=0$. This is because for $q=0.99$ and the assumed temperature the antibonding state participates less in the transport. Furthermore, we show that this picture changes with decreasing  temperature.

\subsection{Case of symmetrical coupling, ($k_{B}T/\Gamma = 0.001$)}
\begin{figure}
\includegraphics[width=\columnwidth]{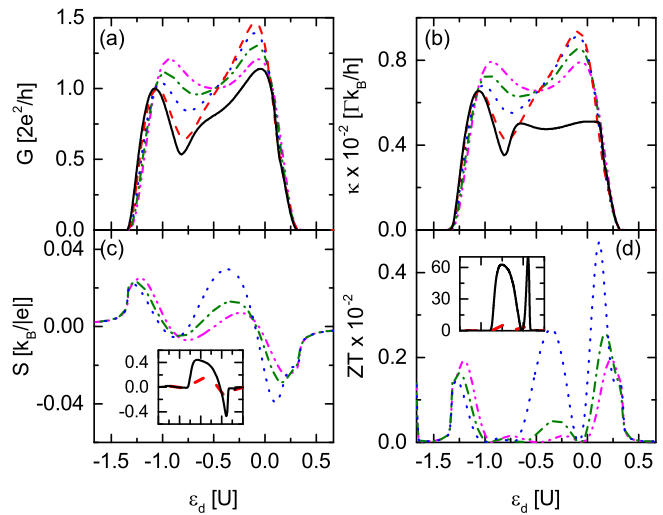}
\caption{\label{Fig:4} Electrical conductance $G$ (a), heat conductance $\kappa$ (b), thermopower $S$ (c), and figure of merit $ZT$ (d), calculated as a function of
the dots' energy level for values of parameter $q$ indicated in the previous figure. The insets show plots of thermopower (c) and figure of merit (d) calculated for $q=0.9$ and $0.99$. The other parameters: $k_{B}T = 0.001\Gamma$, $U=6\Gamma$, $\alpha=1$, $t=1\Gamma$.}
\end{figure}
The picture described above may change dramatically with the change of the temperature. In Fig.~\ref{Fig:4} we show dots level dependence of conductance, heat conductance, thermopower, and corresponding figure of merit calculated for temperature $k_{B}T=0.001\Gamma$. One can notice that the $q=0$ case does not change qualitatively. The quantitative difference can be noticed in Fig.~\ref{Fig:4}(b), where the heat conductance is about ten times smaller than that  shown in Fig.~\ref{Fig:3}(b). With increasing strength of the indirect coupling, $q$, the conductance reveals two peaks. More importantly, the two peak structure survives for $q\approx 1$ in contrast to the situation described in the previous section. Here, even when $q$ is close to 1, no plateau is observed. Here, the antibonding state becomes almost pinned at the Fermi level (as for zero temperature\cite{Vernek}) and retains its position for a broad range of the dots' level positions. As mentioned before, it acquires finite coupling to the leads -- oppositely to the case discussed in Refs.~\onlinecite{Vernek,DingNahm}, and thus, it gives also a contribution to the conductance. It turns out that for lower temperature the intensity of the peak associated with the antibonding state becomes greatly enhanced in comparison with the case discussed above. This suspension of the antibonding peak at the Fermi level and its relatively large intensity results in disappearance of the plateau in the conductance described above. In turn, a two peak structure is present in the conductance. Roughly speaking, one maximum in the conductance appears when the antibonding state is in resonance whereas the second maximum emerges when the peak associated with the bonding state is situated at the Fermi level.
Generally, the dependence of the electron contribution to the heat conductance follows that of the conductance with the exception of the curve calculated for $q=0.99$ where the heat conductance becomes suppressed. The suppression of the heat conductance as $q$ becomes close to 1 can be understood as follows: as $q$ increases, the width of the antibonding peak decreases and therefore the impact to the heat conductance coming from the antibonding resonance also diminishes. The antibonding resonance (for $q$ very close to 1) cuts only a small region of the derivative of the Fermi-Dirac function around the Fermi level. Thus, only the low energetic electrons and holes contribute to the heat conductance. In contrast, for $q \ll 1$, high-energy electrons and holes contribute to transport and lead to increase in $\kappa$.

In Fig.~\ref{Fig:4}(c) the Seebeck coefficient is shown for indicated values of the parameter $q$. With increasing  $q$, the thermopower increases, especially  in the Coulomb blockade regime and when the antibonding level is close to the Fermi level. To understand this feature let us take a closer look at the thermopower in the Coulomb blockade regime. It turns out that for this regime, and for $q$ close to 1, the thermopower is mainly determined by the contribution coming from the antibonding resonance as it is situated in the vicinity of the Fermi level. For $q\approx 1$ the antibonding peak is slightly below the Fermi level of the leads. As described above, with increasing the parameter $q$ the width of the antibonding resonance decreases and for $q$ close to 1 it becomes much smaller than the temperature broadening. As a consequence, the greater part of the antibonding peak lies below the Fermi level (but within temperature broadening $\partial f/ \partial \varepsilon$), which corresponds to hole current greater than electron current, and results in positive thermopower. Conversely, with decreasing the value of the parameter $q$, the peak corresponding to the antibonding state becomes broader, which leads to diminished asymmetry of the peak with respect to the Fermi level, resulting in decreased thermopower. For $q=0$, both bonding and antibonding peaks  are symmetrical around the Fermi level and equally contribute to transport coefficients, thus resulting in symmetric thermopower. Simultaneously, the figure of merit grows non-monotonically with increasing value of the parameter $q$ [Fig.~\ref{Fig:4}(d)].

\subsection{Temperature dependence of transport coefficients}
\label{sec:temp}
\begin{figure}
\includegraphics[width=\columnwidth]{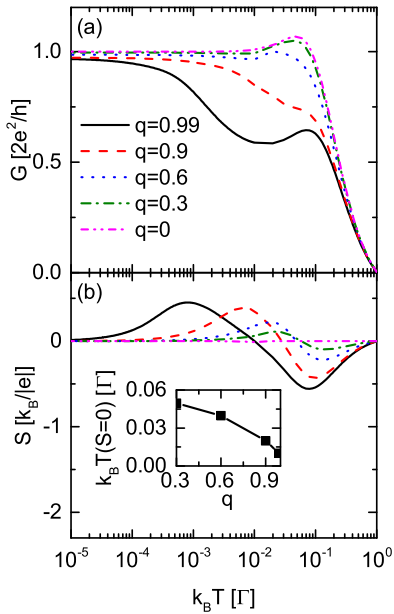}
\caption{\label{Fig:5} Conductance (a) and thermopower (b) as a function of temperature calculated for indicated values of the parameter $q$. The other parameters: $p=0$, $\varepsilon_d=-U/2$, $U=6\Gamma$, $\alpha=1$, $t=1\Gamma$. The inset shows the dependence of temperature for which the Seebeck coefficient vanishes as a function of parameter $q$.}
\end{figure}

To get deeper insight into the influence of temperature on the transport characteristics, we plot temperature dependence of the relevant quantities calculated for $\varepsilon_d=-U/2$. Fig.~\ref{Fig:5} (a) shows temperature dependence of the conductance $G$ and thermopower for indicated values of the parameter $q$. One can notice that for $q=0$ the conductance increases with decreasing temperature achieving a maximum, and then  slightly decreases to the saturation value with a further decrease in the temperature. The conductance remains then constant which results in plateau characteristic for the Kondo effect. The Kondo temperature can be estimated as the temperature at which the conductance reduces
to half of its maximum value. Thus, for $q=0$ one can find $T_K\approx 0.25\Gamma$ and notice that the Kondo temperature decreases with increase in $q$. However, the situation becomes more complex as $q$ tends to unity. One should bear in mind that for $q$ close to unity the bonding and antibonding states in DQD systems become well formed. As mentioned before, the energy levels corresponding to these states differ by widths. More specifically, the bonding (antibonding) level acquires the renormalized width $\tilde{\Gamma}_b=\tilde{\Gamma}(1+q)$ [$\tilde{\Gamma}_a=\tilde{\Gamma}(1-q)$]. As a consequence of the two energy scales, the temperature dependence of the conductance reveals two steps. The first step can be referred to as the activation of the bonding state, whereas the second one can be referred to as the activation of the antibonding state.

Fig.~\ref{Fig:5} (b) shows the temperature dependence of the Seebeck coefficient calculated for $\varepsilon_d=-U/2$ and for indicated values of the parameter $q$. Due to the particle-hole symmetry, the thermopower vanishes for $q=0$ in the whole range of temperature. However, for $q>0$ the density of states ceases to be symmetric with respect to the zero energy, which results in nonzero thermopower. Generally, the absolute value of the thermopower increases with increasing $q$ as the shape of the density of states strongly depends on this parameter. One can also notice, that for specific temperature the Seebeck coefficient vanishes. The point for which the thermopower becomes zero depends on the value of $q$. More specifically, it moves towards lower temperatures with increasing $q$ (see also the inset of Fig.~\ref{Fig:5} (b)). Thus, this feature can indirectly give us some information about the behavior of the Kondo temperature. Generally, one can notice that the thermoelectric transport beyond the temperature for which $S=0$ is hole-like, whereas above this temperature the electrons become majority carriers.

\begin{figure}[t]
\includegraphics[width=\columnwidth]{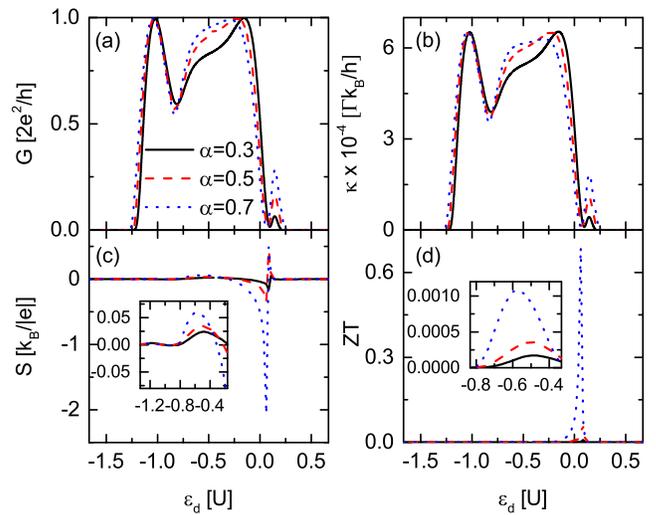}
\caption{\label{Fig:6}Electrical conductance $G$ (a), heat conductance $\kappa$ (b), thermopower $S$ (c), and figure of merit $ZT$ (d), calculated as a function of
the dots' energy level for indicated values of parameter $\alpha$. The insets show a magnification of thermopower (c) and figure of merit (d). The other parameters: $k_{B}T =10^{-4}\Gamma$, $U=6\Gamma$, $q=1$, $t=1\Gamma$.}
\end{figure}

\subsection{Asymmetrical coupling}
Let us now consider the influence of asymmetry in coupling of the two dots to a given lead, measured by the parameter $\alpha$. In Fig.~\ref{Fig:6} the dots' level dependence of basic thermoelectric coefficients is displayed for indicated values of the parameter $\alpha$. For $\alpha\in (0,1)$, the conductance [Fig.~\ref{Fig:6}(a)] and electron contribution to the heat conductance [Fig.~\ref{Fig:6}(b)] reveal behavior typical for the Fano effect\cite{Fano,TrochaFano}. Due to destructive quantum interference, the antiresonance structure appears in the conductance.
With the increase of the parameter $\alpha$, both conductances increase for dots' energy level near the particle-hole symmetry point and for a narrow range of energy above $\varepsilon_{d}=0$, where the transport through antibonding state is dominant. This effect leads also to pronounced thermopower in the vicinity of the point for which the antiresonance occurs [see Fig.~\ref{Fig:6}(c)].

The increase in the parameter $\alpha$ greatly influences the thermoelectric effects due to the fact that the slope of the antiresonance dip in transmission function changes. The higher the $\alpha$, the steeper is the antipeak, up to a point where the Fano effect disappears and there is only one broad peak corresponding to the superradiant state as described in Sec. IIIA. In the particle-hole symmetry point, however, the thermopower is two orders of magnitude smaller than in the vicinity of the antiresonance, as shown in the inset to Fig.~\ref{Fig:6}(c). Moreover, there is a difference in the slope of thermopower, which is positive near the antiresonance and negative around the particle-hole symmetry point. The sign of the Seebeck coefficient around the antiresonance also indicates that the main carriers are electrons, while hole-like transport is dominant in the deep Kondo regime.

The suppression of thermal conductance and enhancement of the Seebeck coefficient results in relatively large magnification of the figure of merit for the dots' level position corresponding to the antiresonance [Fig.~\ref{Fig:6}(d)]. There is also an increase near the particle-hole symmetry point as shown in the inset to Fig.~\ref{Fig:6}(d).

\section{Magnetic leads: absence of spin accumulation}
In this section we consider the DQD system when both electrodes
are ferromagnetic, with the corresponding spin polarization factor $p$.
In the following considerations only collinear, i.e., parallel
and antiparallel, configurations will be analyzed.

\subsection{Exchange field}
\begin{figure}[t]
\includegraphics[width=\columnwidth]{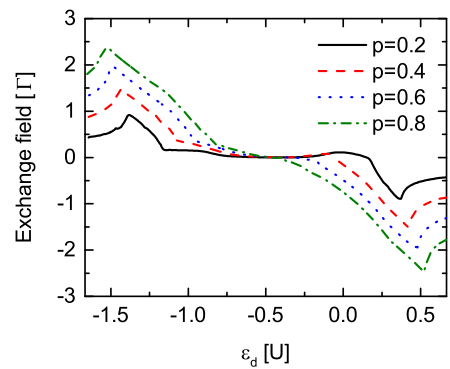}
\caption{\label{Fig:7}Exchange field as a function of the dots' energy level for indicated values of polarization in parallel configuration. The other parameters: $k_{B}T = 0.01\Gamma$, $U=6\Gamma$, $q=0.99$, $\alpha=1$, $t=1\Gamma$.}
\end{figure}
It is a well-known fact that for quantum dot systems strongly coupled to ferromagnetic electrodes, the energy levels become split due to the exchange field generated by difference in couplings to the minority and majority carriers in the leads.~\cite{Martinek,karwackiJPCM,Wojcik2} However, this effect is of importance only for the parallel magnetic configuration of the leads. In turn, for the antiparallel configuration, no exchange field appears as the fields originating from both leads become compensated for the symmetric system, i.e., $\mathbf{\Gamma}_{L\sigma}=\mathbf{\Gamma}_{R\sigma}$. Within the SBMF technique the exchange field, renormalizing dots' energy levels $\tilde{\varepsilon}_{\sigma}$ for $\sigma=\uparrow,\downarrow$ can be expressed as a difference between the energy levels $\tilde{\varepsilon}_{\uparrow}-\tilde{\varepsilon}_{\downarrow}$, which simplifies to the difference between the appropriate spin-dependent Lagrange multipliers, $\lambda_{\uparrow}^{(2)}-\lambda_{\downarrow}^{(2)}$. In Figure~\ref{Fig:7} we present the exchange field as a function of the dots' level energy calculated for indicated values of the leads' polarization $p$. The exchange field vanishes in the particle-hole symmetric point $\varepsilon_{d}=-U/2$, but increases (regarding its absolute value) away from this point. A significant increase is noticed in the empty orbital regime, i.e., for energies above $\varepsilon_{d}=0$ and in the twofold occupancy regime, i.e., below $\varepsilon_{d}=-U$. Furthermore, the magnitude of this field increases with the increase in polarization of the leads.
\begin{figure}[t]
\includegraphics[width=\columnwidth]{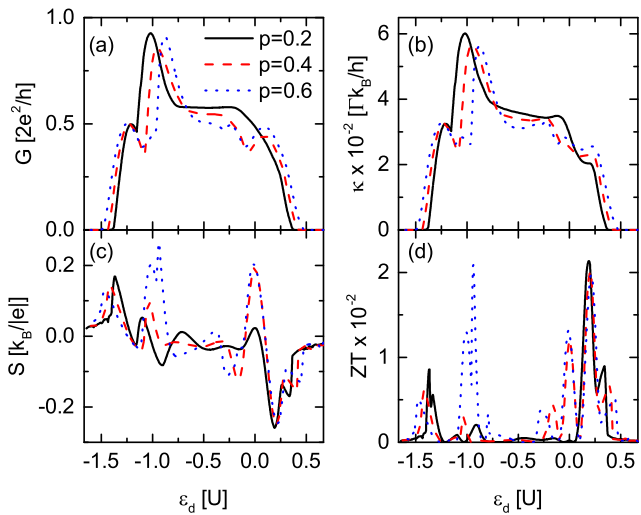}
\caption{\label{Fig:8}Electrical conductance $G$ (a), heat conductance $\kappa$ (b), thermopower $S$ (c), and figure of merit $ZT$ (d), calculated as a function of
the dots' energy level for indicated values of polarization $p$ and parallel configuration. The other parameters: $k_{B}T = 0.01\Gamma$, $U=6\Gamma$, $q=0.99$, $\alpha=1$, $t=1\Gamma$.}
\end{figure}

\subsection{Effect of exchange field on transport properties}
Existence of an exchange field significantly modifies transport properties of the considered system. Figure~\ref{Fig:8} (a) presents electrical conductance as a function of the energy level for indicated values of polarization.
The conductance associated with the bonding peak near $\varepsilon_{d}=-U$ becomes split. Similarly, near $\varepsilon_{d}=0$ a small kink develops. With increasing parameter $p$, these features become more pronounced. Moreover, for high magnetic polarization, the flat region in the conductance corresponding to the singly occupied regime vanishes.

The exchange field has also significant impact on the thermoelectric properties of the considered system. Firstly, due to splitting of the density-of-states peaks the thermopower reveals more points at which the Seebeck coefficient is zero. Moreover, the increase in the magnetic polarization of the external electrodes leads to enhancement of the thermopower in the vicinity of $\varepsilon_d=0$. Further increase in polarization $p$ causes relatively large increase in $S$ near $\varepsilon_d=-U$ [see the curve for $p=0.6$ in Fig.~\ref{Fig:8} (c)].

On the other hand, the thermoelectric efficiency of the system, measured by figure of merit $ZT$ shown in Fig.~\ref{Fig:8} (d), stays roughly the same for indicated values of the parameter $p$. However, for sufficiently large $p$, a pronounced value of $ZT$ can be found for the dots' level positions close to $\varepsilon_d=-U$. This result is in contrast with the one obtained for the DQD system for temperatures above the Kondo temperature, where the figure of merit monotonically decreases with increasing $p$\cite{Trocha-thermo}.

\begin{figure}[t]
\includegraphics[width=\columnwidth]{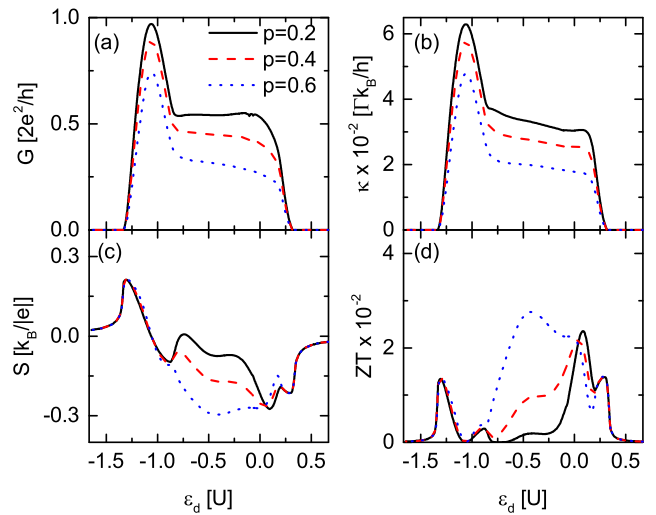}
\caption{\label{Fig:9}Electrical conductance $G$ (a), heat conductance $\kappa$ (b), thermopower $S$ (c), and figure of merit $ZT$ (d), calculated as a function of
the dots' energy level for indicated values of polarization $p$ and anti-parallel configuration. The other parameters: $k_{B}T = 0.01\Gamma$, $U=6\Gamma$, $q=0.99$, $\alpha=1$, $t=1\Gamma$.}
\end{figure}
As mentioned before, the exchange field does not play any role in the antiparallel configuration for symmetric systems. In the antiparallel alignment, conductance [Fig.~\ref{Fig:9} (a)] and electron contribution to the heat conductance [Fig.~\ref{Fig:9} (b)] are similar to the corresponding quantities calculated for nonmagnetic systems. However, with increasing polarization, the conductance diminishes. This is due to the {\it bottleneck} effect occurring in the antiparallel configuration for both spin channels as the polarization increases.
The thermopower becomes influenced by the leads' polarization for the antiparallel configuration as well. Specifically, the $S$ increases in the single-occupied regime with increasing $p$. Accordingly, this increase in thermopower leads to pronounced $ZT$ in the Coulomb blockade regime.

\section{Magnetic leads: case of spin accumulation}
While in the previous section a spin-dependent transport was discussed, a condition of relatively small spin relaxation in the electrodes may lead to spin accumulation and therefore to the rise of the spin thermoelectric effect. Here we consider spin counterparts to charge thermopower and figure of merit.

\subsection{Case of symmetrical coupling}
\begin{figure}
\includegraphics[width=\columnwidth]{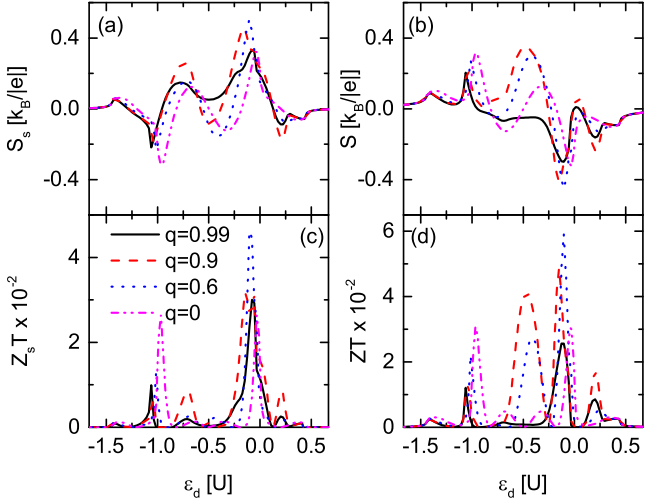}
\caption{\label{Fig:10}Spin and charge Seebeck coefficients [(a) and (b)] and spin and charge figure of merit [(c) and (d)] for the case of spin accumulation in the system as a function of dots' energy level and indicated values of the parameter $q$. The other parameters: $k_{B}T=0.01\Gamma$, $p=0.4$, $\alpha=1$, $t=1\Gamma$, $U=6\Gamma$.}
\end{figure}

\begin{figure}
\includegraphics[width=\columnwidth]{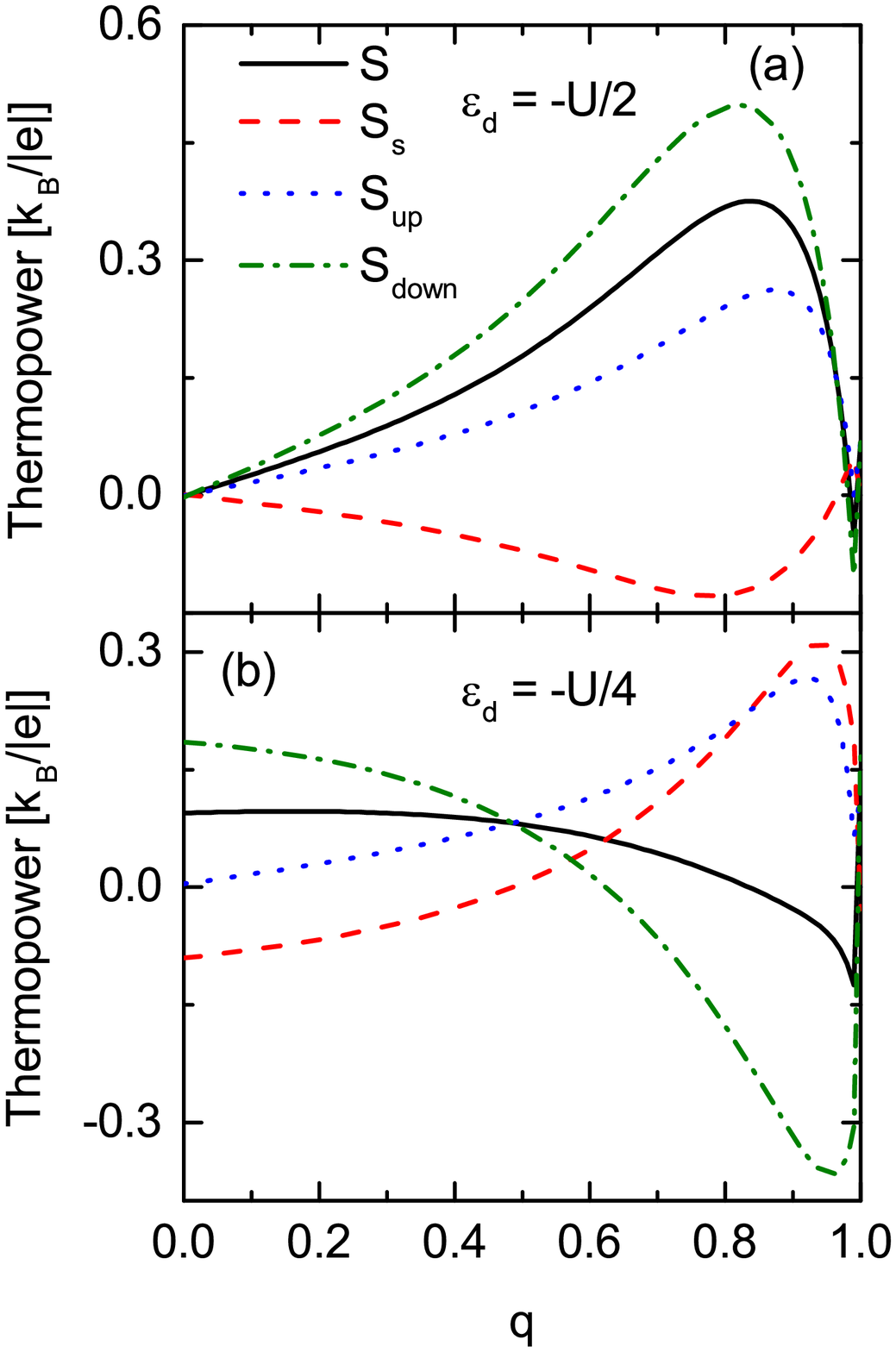}
\caption{\label{Fig:11} Spin and charge Seebeck coefficients along with spin-up and spin-down contributions as a function of parameter $q$ for $\varepsilon_{d}=-U/2$ (a) and $\varepsilon_{d}=-U/4$ (b). The other parameters: $k_{B}T=0.01\Gamma$, $p=0.4$, $\alpha=1$, $t=1\Gamma$, $U=6\Gamma$.}
\end{figure}

\begin{figure}
\includegraphics[width=\columnwidth]{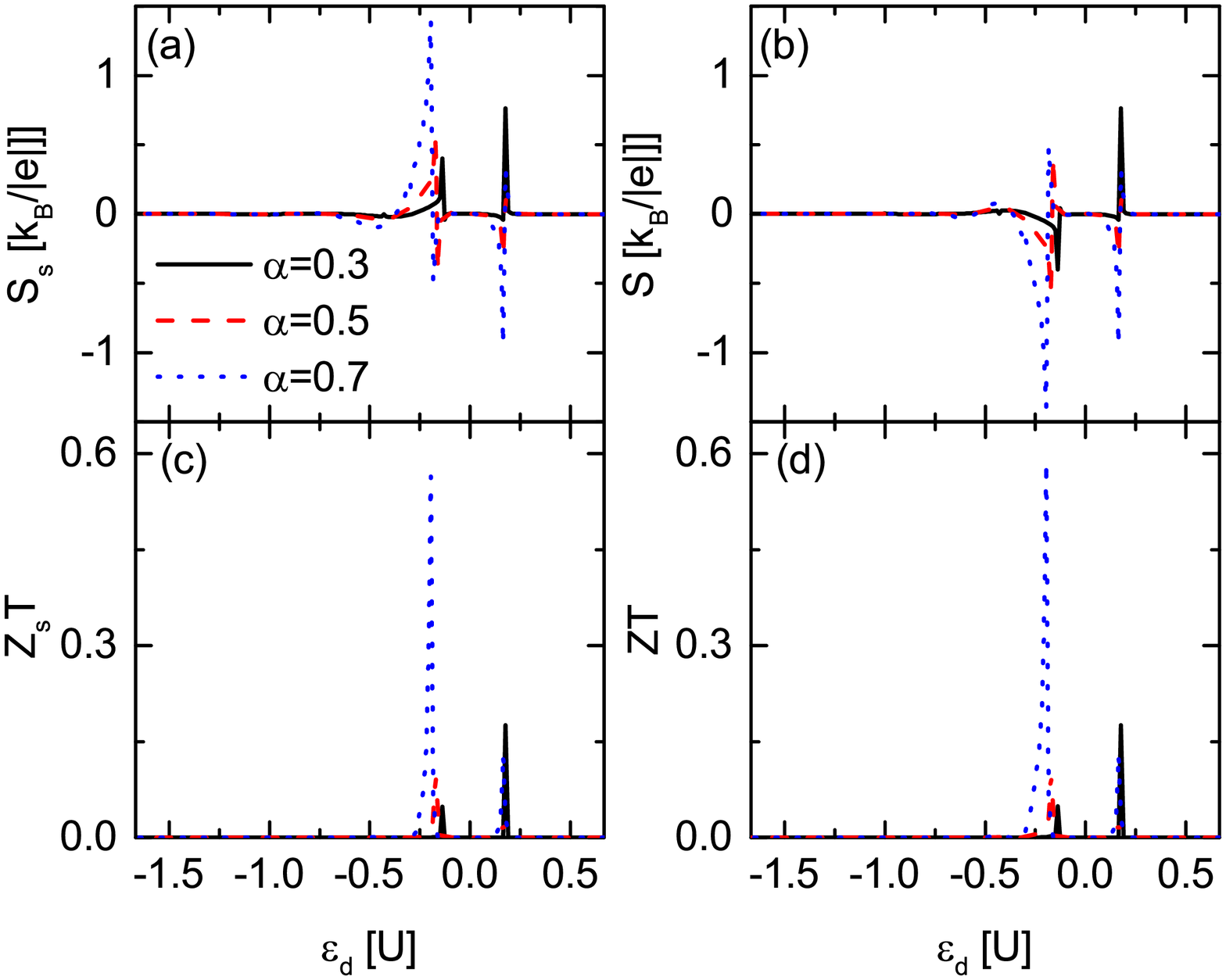}
\caption{\label{Fig:12}Spin and charge Seebeck coefficients [(a) and (b)] and spin and charge figure of merit [(c) and (d)] for the case of spin accumulation in the system as a function of dots' energy and indicated values of the parameter $\alpha$. The other parameters: $k_{B}T=10^{-4}\Gamma$, $p=0.4$, $q=1$, $t=1\Gamma$, $U=6\Gamma$.}
\end{figure}

\begin{figure}
\includegraphics[width=\columnwidth]{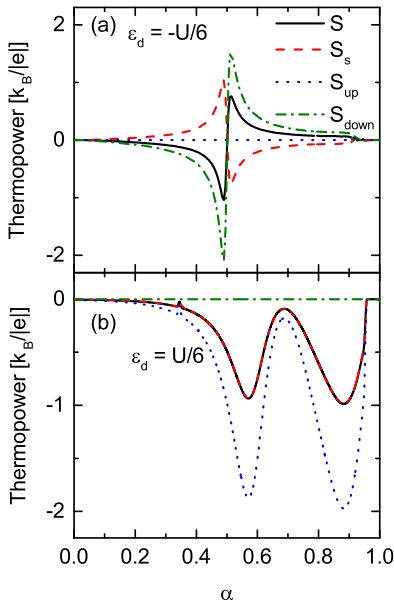}
\caption{\label{Fig:13} Spin and charge Seebeck coefficients along with spin-up and spin-down contributions as a function of parameter $\alpha$ for $\varepsilon_{d}=-U/6$ (a) and $\varepsilon_{d}=U/6$ (b). The other parameters: $k_{B}T=10^{-4}\Gamma$, $p=0.4$, $q=1$, $t=1\Gamma$, $U=6\Gamma$.}
\end{figure}

Fig.~\ref{Fig:10} presents spin and charge Seebeck coefficients along with respective spin and charge figure of merit coefficients as a function of dots' energy level and indicated values of the parameter $q$. Fig.~\ref{Fig:10} (a) and (b) present spin and charge Seebeck coefficients respectively. Similarly to the previously discussed cases, for $q=0$ the particle-hole symmetry is preserved, which is evidenced by zero value of both coefficients in the particle-hole symmetry point (i.e., $\varepsilon_{d}=-U/2$). It also means that the net spin current is zero due to the fact that spin-up and spin-down components to the current suppress each other.

Additionally, the figures of merit visualized in Fig.~\ref{Fig:10}(c) and (d) present similar symmetry and become zero in the particle-hole symmetry point as well. However, the increase in the parameter $q$ leads to the breaking of the symmetry. All the parameters change non-monotonically, increasing with $q$ up to $q=0.9$, then decreasing drastically for $q>0.9$. The results indicate that the charge thermopower is more sensitive to changes in the parameter $q$, which is suggested by strong deviation from zero value of the thermopower near $\varepsilon_{d}=-U/2$. This fact can be explained by considering spin-up and spin-down contributions to the thermopower presented in Fig.~\ref{Fig:11}(a) along with spin and charge Seebeck coefficients for $\varepsilon_{d}=-U/2$. Here, as previously stated, for $q=0$ all Seebeck coefficients are zero. Increase in $q$ leads to the increase in the spin-up and down contributions to the thermopower, and thus to increase in absolute values of the charge and spin thermopowers. It is noteworthy, that for $q\approx 0.9$ the absolute values of Seebeck coefficients assume maximal values, while further increase in $q$ leads to dramatic decline. For $\varepsilon_{d}=-U/4$ presented in Fig.~\ref{Fig:11}(b), the spin-down contribution to the thermopower changes sign for $q\approx 0.5$, while spin-up contribution retains its sign over the whole range of parameter $q$. This fact, however, greatly improves the value of spin thermopower. Furthermore, this increase results in a larger spin figure of merit, as shown in Fig.~\ref{Fig:10}(c). However, due to the aforementioned symmetry breaking, the charge figure of merit in Fig.~\ref{Fig:10}(d) assumes larger values than its spin counterpart.

\subsection{Case of asymmetrical coupling}
Fig.~\ref{Fig:12} presents spin and charge Seebeck coefficients and spin and charge figure of merit coefficients accordingly as a function of dots' energy level for different values of parameter $\alpha$. Thermoelectric coefficients are significant in two particular values of the dots' energy level $\varepsilon_{d}\approx\pm \tilde{t}$, where they assume their maximal values due to emergence of the bonding and antibonding states. Increasing the parameter $\alpha$ results in much greater spin thermopower for $\varepsilon_{d}\approx-\tilde{t}$ than for $\varepsilon_{d}\approx\tilde{t}$. This fact is especially noticeable in the plot of the spin figure of merit in Fig.~\ref{Fig:12}(c), where the thermoelectric efficiency increases significantly for $\alpha>0.5$, while it remains constant for $\varepsilon_{d}\approx\tilde{t}$.

Fig.~\ref{Fig:12}(b) and (d) present charge thermopower and figure of merit accordingly. Behavior of those coefficients is similar to their spin counterparts, with the exception of charge thermopower for $\varepsilon_{d}\approx-\tilde{t}$, where it is of the opposite sign to spin thermopower. For $\varepsilon_{d}\approx\tilde{t}$, however, the sign of spin and charge Seebeck coefficients is the same for all values of $\alpha$, and the values of both Seebeck coefficients are comparable. This is a result of the fact, that in the vicinity of $\varepsilon_{d}\approx\tilde{t}$, the  spin-down contribution to thermopower is close to zero, while the spin-up contribution is finite and greater than zero. The opposite is true for $\varepsilon_{d}\approx-\tilde{t}$, where the spin-up contribution is approximately zero while the spin-down contribution is finite and greater than zero. For other values of the dots' energy level, both contributions are negligible. This fact is shown in Fig.~\ref{Fig:13}, where both contributions and spin and charge Seebeck coefficients are plotted as a function of the parameter $\alpha$.

In Fig.~\ref{Fig:13}(a) the spin-up contribution is approximately zero for the whole range of the parameter $\alpha$, whereas spin-down contribution is finite. This explains different signs of spin and charge thermopower in Fig.~\ref{Fig:12}(a) and (b). Additionally, the spin-down contribution to thermopower changes its sign for $\alpha=0.5$ and decreases (with respect to its absolute value) for $\alpha\neq 0.5$. In the case of $\varepsilon_{d}\approx\tilde{t}$, presented in Fig.~\ref{Fig:13}(b), the spin-down contribution to the thermopower is negligible compared to the spin-up contribution in the whole range of the  parameter $\alpha$. The spin-up component, however, exhibits two peaks for $\alpha\approx 0.6$ and $0.9$.

\section{Conclusions}
In conclusion, we have shown that signatures of quantum interference in strongly correlated double quantum dot systems can be visible in the thermoelectric effects for sufficiently low temperatures. Moreover, they can strongly modify and enhance the thermoelectric effects.

These signatures take the form of an asymmetry in the conductance, reminiscent of the Dicke (linewidth narrowing of the peak corresponding to the antibonding / subradiant state) and Fano (antiresonance in the vicinity of the antibonding state) effects. This asymmetry has also significant impact on the  thermopower and figure of merit, leading to an increase in those parameters.

Similar discussion has been also presented for spin-dependent thermoelectric effects in the case of ferromagnetic leads. Here, an additional structure in the peaks appears due to the exchange field resulting from the ferromagnetic leads in the parallel configuration. In the antiparallel configuration, in turn,  increasing  spin polarization of the leads suppresses electronic transport, but can lead to an increase in the thermopower and the thermoelectric figure of merit.
When the spin accumulation in the leads is significant, a spin thermoelectric effect can be observed. By an appropriate combination of the parameters $q$ and $\alpha$, closely related to interference effects in the considered system, one can improve the spin thermoelectricity.

\begin{acknowledgments}
\L{}.K. acknowledges support from the National Science Centre in Poland under Project No. DEC-2012/04/A/ST3/00372. The authors would like to thank J\'{o}zef Barna\'{s} and Ireneusz Weymann for carefully reading the manuscript and providing helpful comments.
\end{acknowledgments}

\end{document}